\documentclass[doublecol]{epl2}

\usepackage{amsmath}
\usepackage{amssymb}
\usepackage{bm}
\usepackage{graphics}
\usepackage{float}
\usepackage{slashed}

\title{Experimental signatures of a new dark matter WIMP}

\author{Reagan Thornberry, Maxwell Throm, Gabriel Frohaug, John Killough, Dylan Blend, \\
Michael Erickson, Brian Sun, Brett Bays, and Roland E. Allen}

\shortauthor{Thornberry, Throm, Frohaug, Killough, Blend, Erickson, Sun, Bays, and Allen}

\institute{Physics and Astronomy, Texas A\&M University, College Station, Texas 77843, USA}


\abstract{The WIMP proposed here yields the observed abundance of dark matter, and is consistent with the current limits from direct detection, indirect detection, and collider experiments, if its mass is $\sim 72$ GeV/$c^2$. It is also consistent with analyses of the gamma rays observed by Fermi-LAT from the Galactic center (and other sources), and of the antiprotons observed by AMS-02, in which the excesses are attributed to dark matter annihilation. These successes are shared by the inert doublet model (IDM), but the phenomenology is very different: The dark matter candidate of the IDM has first-order gauge couplings to other new particles, whereas the present candidate does not. In addition to indirect detection through annihilation products, it appears that the present particle can be observed in the most sensitive direct-detection and collider experiments currently being planned.}

\begin{document}

\maketitle

As is well--known, there are a vast number of hypothetical dark matter
candidates, most of which do not have well-defined masses or couplings, and
many of which have already been ruled out by experiment. The most popular
single candidate has been 
the lightest neutralino $\chi ^{0}$ of supersymmetry (susy)~\cite{susy-DM-1996}.
 But, as is also well-known, faith in this candidate has
diminished during the past few years, because neither a susy dark matter
WIMP nor other susy particles have been observed, despite strenuous efforts.
In addition, the simplest susy models which have ``natural'' values for the
parameters, and which are also compatible with limits from the LHC, are
found to be in disagreement with both the abundance of dark matter and the
limits from direct-detection 
experiments~\cite{Baer-Barger-2016,Baer-Barger-2018,Roszkowski-2018,Baer-Barger-2020,Tata-2020}.

Here we propose an alternative candidate which in some respects resembles
the neutralino, but which would open up a whole new sector with new
particles and new physics which can be observed in the foreseeable future.
We have named particles of this new kind ``higgsons''~\cite{statistical,DM1,DM2,DM3
}, 
represented by $H$, to distinguish them
from Higgs bosons $h$ and the higgsinos $\widetilde{h}$ of susy. The
lightest neutral particles in these three groups are $H^{0}$, $h^{0}$, and 
$\widetilde{h}^{0}$. We will demonstrate below that these new particles
(higgsons) have very favorable features -- after the theory has been
reformulated in the way described below.

Many of these features are shared by the inert doublet model (IDM) -- introduced
long ago~\cite{Ma-1978} and later proposed as an explanation of dark matter~\cite{Ma-2006} 
-- for which the phenomenology has been very extensively explored
in many papers. Only a representative sample can be cited 
here~\cite{Barbieri-2006,Honorez-2007,Lundstrom,Miao,Honorez-2010,Gustafsson-2012,Klasen-2013,Goudelis-2013,Eiteneuer,Belyaev-2018,Dutta,Dercks-2019,Robens,Belyaev-2019,Banerjee},
but the basic
idea is that the Standard Model Higgs doublet is supplemented by a second
doublet of scalar fields which have odd parity under a postulated new $Z_{2}$
symmetry. To avoid confusion, these fields will be distinguished by a
subscript $I$: 
\begin{align}
\left( 
\begin{array}{c}
H_{I}^{+} \\ 
\frac{1}{\sqrt{2}} \left( H_{I}^{0}+ i A_{I}^{0} \right)
\end{array}
\right) \;.
\end{align}
Since each of these fields is odd with respect to the hypothetical $Z_{2}$,
whereas all other fields are even, every term in the action must involve an
even number of these fields. This makes the dark matter candidate $H_{I}^{0}$
stable, because it has the lowest mass. In addition, many of the successes 
described below for the present candidate, called $H^{0}$ here, are shared by 
$H_{I}^{0}$. 

However, in the phenomenology of the IDM there are
numerous important processes that involve first-order gauge
couplings of $H_{I}^{0}$ to the other particles $A_{I}^{0}$ and $H_{I}^{\pm }$.
See, e.g., Figs. 3 and 4 of \cite{Honorez-2007}; Fig.~2 of \cite{Lundstrom}; Fig.~1 of \cite{Miao}; Fig~4 of \cite{Gustafsson-2012}; 
Figs.~9, 10, 11, 13, 14, 16, 17, 18, and 21 
of \cite{Belyaev-2018}; Figs.~2 and 3 of \cite{Robens}; 
Fig.~6   of \cite{Belyaev-2019}; Figs.~1 and 3 of \cite{Banerjee}; 
and the constraints on the IDM parameters in Section 4.1 of 
\cite{Eiteneuer}, which imply that the masses of $A_{I}^{0}$ 
and $H_{I}^{\pm }$ should be reasonably near that of $H_{I}^{0}$.

 As will be seen below, the present dark matter
candidate $H^{0}$ has only second-order gauge couplings -- and these are to
only itself plus the gauge bosons. This results in a very different
phenomenology, with a substantial reduction in the number of observable 
processes involved in annihilation, scattering, and creation. There are
other differences: $H^{0}$ is its own antiparticle, it does not require an
extra $Z_{2}$ symmetry, and it is related to Higgs bosons in a very
different way. 

We begin with the primitive 2-component spin 1/2 bosonic fields $\Phi _{r}$
introduced in our previous papers, which are joined in an $n$-component
gauge multiplet $\widetilde{\Phi }_{r}$. (We name each multiplet 
$\widetilde{\Phi }_{r}$ after a typical member $\Phi _{r}$ in order to simplify the
notation.) Each $\Phi _{r}$ transforms as a 2-component spinor under
rotations, but is left unchanged by a boost.

It will be seen that the action for the physical fields $\Phi _{R}$, as
defined below, is invariant under both rotations and boosts. The final
theory is in fact fully Lorentz invariant.

For simplicity, we begin with a single (e.g. grand-unified) gauge field
having covariant derivative 
\begin{align}
D_{\mu }=\partial _{\mu }-iA_{\mu }\quad ,\quad A_{\mu }=A_{\mu }^{j}t^{j}
\end{align}
(so that the coupling constant is temporarily absorbed into $A_{\mu }$) and
field strength 
\begin{align}
\hspace{-0.2in}\quad F_{\mu \nu }=F_{\mu \nu }^{j}t^{j}\quad ,\quad F_{\mu
\nu }^{j}=\partial _{\mu }A_{\nu }^{j}-\partial _{\nu }A_{\mu
}^{j}+c_{ii^{\prime }}^{j}A_{\mu }^{i}A_{\nu }^{i^{\prime }}\;.
\label{eq3}
\end{align}
Later we will specialize to the electroweak theory after symmetry breaking.
Natural units $\hbar =c=1$ are used, with the $(-+++)$ convention for the
metric tensor. Summations are never implied over a repeated gauge index like $r$
or $R$, but are always implied over repeated coordinate indices like $\mu
,\nu =0,1,2,3$ and $k=1,2,3$, as well as the index $i$ or $j$ labeling gauge
generators $t^{j}$. The same name and symbol are used for a field and the
particle which is an excitation of that field.

The initial action for the new fields $\widetilde{\Phi }_{r}$ is 
\begin{align}
S_r = \int d^{4}x\, \widetilde{\Phi }_{r}^{\dag }\left( x\right) & D^{\mu
}D_{\mu }\, \widetilde{\Phi }_{r}\left( x\right)  \notag \\
& +\int d^{4}x\,\widetilde{\Phi }_{r}^{\dag }\left( x\right) \,
\overrightarrow{B}\left( x\right) \cdot \overrightarrow{\sigma }
\;\widetilde{\Phi }_{r}\left( x\right)  \label{e1}
\end{align}
where the 3-vectors $\overrightarrow{\sigma }$ and $\overrightarrow{B}$
respectively contain the Pauli matrices $\sigma^k$ and the ``magnetic''
components of the field strength tensor $F_{\mu \nu }$: 
\begin{align}
F_{kk^{\prime }}=-\varepsilon _{kk^{\prime }k^{\prime \prime }}B_{k^{\prime
\prime }}\;.  \label{e7}
\end{align}
(This action is derived in Ref.~\cite{statistical} but here is taken to be a
phenomenological postulate~\cite{DM1}.) At this point we deviate from our
earlier papers by requiring all fundamental physical fields $\widetilde{\Phi 
}_{R}$ to satisfy Lorentz invariance. This means that the anomalous second
term in (\ref{e1}) must disappear from the action. This can be achieved by
requiring the physical fields to assume one of the two forms defined below,
which are respectively called \textit{Higgs/amplitude fields} and 
\textit{higgson fields}.

\textbf{Higgs/amplitude fields} are defined to be those for which 
\begin{align}
\widetilde{\Phi }_{R}^{\dag }\left( x\right) \,\overrightarrow{\sigma }\; 
\widetilde{\Phi }_{R}\left( x\right) =0
\end{align}
(where $\overrightarrow{\sigma }$ always implicitly multiplies an
appropriate identity matrix). These fields -- analogous to the
Higgs/amplitude modes observed in superconductors~\cite{Varma} -- are
obtained by writing the $4n$-component $\widetilde{\Phi }_{R}$ in terms of $2n$-component 
fields $\widetilde{\Phi }_{r}$ and $\widetilde{\Phi }_{r^{\prime }}$ 
with opposite spins (and equal amplitudes): 
\begin{align}
\widetilde{\Phi }_{R}=\left( 
\begin{array}{c}
\widetilde{\Phi }_{r} \\ 
\widetilde{\Phi }_{r^{\prime }}
\end{array}
\right)  \label{e2}
\end{align}
so that 
\begin{align}
\widetilde{\Phi }_{R}^{\dag }\left( x\right) \,\overrightarrow{\sigma }\, 
\widetilde{\Phi }_{R}\left( x\right) &=\left( 
\begin{array}{cc}
\widetilde{\Phi }_{r}^{\dag } & \widetilde{\Phi }_{r^{\prime }}^{\dag }
\end{array}
\right) \left( 
\begin{array}{cc}
\overrightarrow{\sigma }\, & 0 \\ 
0 & \overrightarrow{\sigma }\,
\end{array}
\right) \left( 
\begin{array}{c}
\widetilde{\Phi }_{r} \\ 
\widetilde{\Phi }_{r^{\prime }}
\end{array}
\right)  \notag \\
&=\widetilde{\Phi }_{r}^{\dag }\,\overrightarrow{\sigma }\,\widetilde{\Phi }
_{r}+\widetilde{\Phi }_{r^{\prime }}^{\dag }\,\overrightarrow{\sigma }\, 
\widetilde{\Phi }_{r^{\prime }}  \label{eq103a} \\
&=0\;.
\end{align}
In this case we write for each gauge component 
\begin{align}
\Phi _{R}\left( x\right) =\phi _{R}\left( x\right) \,\xi _{R}\quad \mathrm{\
with}\quad \xi _{R}^{\,\dag }\,\xi _{R}=1
\end{align}
where $\xi _{R}$ has $4$ constant components and $\phi _{R}\left( x\right) $
is a $1$-component complex amplitude. Then (\ref{e1}) and its counterpart
for $r^{\prime }$ reduce to 
\begin{align}
S_{R}=\int d^{4}x \, \mathcal{L}_{R}\quad ,\quad \mathcal{L}_{R}=
\widetilde{\phi }_{R}^{\dag }\left( x\right) D^{\mu }D_{\mu }\,\widetilde{\phi }
_{R}\left( x\right)
\end{align}
where $\widetilde{\phi }_{R}$ contains all the gauge components $\phi _{R}$.

In the electroweak theory, we interpret $\widetilde{\phi }_{R}$, with
components $\phi _{R}^{0}$ and $\phi _{R}^{+}$, as the usual Higgs doublet
(in the Higgs basis), with $\phi _{R}^{0}$ condensing and supporting Higgs
boson excitations $h_{R}^{0}$: 
\begin{align}
\phi _{R}^{0}=v_{R}^{0}+h_{R}^{0}\quad ,\quad v_{R}^{0}=\langle \phi
_{R}^{0}\rangle \;.
\end{align}
We have assumed a pair of bosonic doublets $\widetilde{\Phi }_{r}$ and 
$\widetilde{\Phi }_{r^{\prime }}$ with the same gauge quantum numbers --
i.e., (weak) isospin and hypercharge -- in somewhat the same spirit as in
standard susy. We choose the convention that $\widetilde{\Phi }_{r}$ has
spin up and $\widetilde{\Phi }_{r^{\prime }}$ spin down for the field 
$\widetilde{\Phi }_{R}$ of (\ref{e2}) which condenses. There is then another
independent field $\widetilde{\Phi }_{R^{\prime }}$ in which $\widetilde{\Phi }_{r}$ 
has spin down and $\widetilde{\Phi }_{r^{\prime }}$ spin up.
This field has no condensate. Only the kinetic term of (\ref{e1}) has been
treated at this point, without the potentially quite complicated terms that
yield masses and further interactions.

\textbf{Higgson fields} are defined to be those for which 
\begin{align}
\Phi _{S}^{\dag }\left( x\right) \,\overrightarrow{B}\;\Phi _{S}\left( x\right) =0
\end{align}
(where $\overrightarrow{B}$ implicitly multiplies a $4\times 4$ identity
matrix, and the subscripts $s$ and $S$ will be used in this context to avoid confusion). 
This can be achieved by writing the $4$-component $\Phi _{S}$ in terms
of a $2$-component field $\Phi _{s}$ and its charge conjugate $\Phi _{s}^{c}$: 
\begin{align}
\Phi _{S}=\frac{1}{\sqrt{2}}\left( 
\begin{array}{c}
\Phi _{s} \\ 
\Phi _{s}^{c}
\end{array}
\right) \;.
\end{align}

From (\ref{eq3}) and (\ref{e7}) we have 
\begin{align}
B_{k^{\prime \prime }}=-\varepsilon _{k^{\prime \prime }kk^{\prime
}}F_{kk^{\prime }}\quad ,\quad F_{kk^{\prime }}=F_{kk^{\prime }}^{j}t^{j}\;.
\end{align}
Also, $\Phi _{s}$ and $\Phi _{s}^{c}$ have opposite gauge quantum numbers --
i.e. opposite expectation values for the generators $t^{j}$ (which are here treated
as operators rather than matrices):
\begin{align}
\Phi _{s}^{c\,\dag }t^{j}\,\Phi _{s}^{c}=-\,\Phi _{s}^{\dag }t^{j}\,\Phi
_{s}\;.
\end{align}
It follows that 
\begin{align}
\Phi _{S}^{\dag }\left( x\right) \,\overrightarrow{B}\,\Phi _{S}\left(
x\right) & =\left( 
\begin{array}{cc}
\Phi _{s}^{\dag } & \Phi _{s}^{c\,\dag }
\end{array}
\right) \left( 
\begin{array}{cc}
\overrightarrow{B}\, & 0 \\ 
0 & \overrightarrow{B}\,
\end{array}
\right) \left( 
\begin{array}{c}
\Phi _{s} \\ 
\Phi _{s}^{c}
\end{array}
\right)  \\
& =\Phi _{s}^{\dag }\,\overrightarrow{B}\,\Phi _{s}+\Phi _{s}^{c\,\dag }\,
\overrightarrow{B}\,\Phi _{s}^{c} \\
& =0\;.
\end{align}
We have thus satisfied relativistic invariance by introducing the bosonic
analog of Majorana fields.

At this point three comments are appropriate: (1)~All the higgson fields 
$\Phi _{S}$ are charge-neutral. (2)~In the following we will consider only the 
$\Phi _{S}$ with no condensate. (3) The lowest-mass 
excitation of such a field, called $H^{0}$ below, is not the only stable
higgson, but it will emerge from the early universe with the highest
density: The more massive particles will fall out of equilibrium
earlier and be rapidly thinned out by the subsequent expansion, and they also
will have larger annihilation cross-sections.

Before the addition of mass and further interaction terms, the action for
any higgson field is
\begin{align}
\int d^{4}x\,\Phi _{S}^{\dag }\,D^{\mu }D_{\mu }\,\Phi _{S}\;.
\end{align}
(This follows from the invariance of the first term in (\ref{e1}) under charge
conjugation.) The mass eigenstates will then have an action 
\begin{align}
S_{S}& =\int d^{4}x\,\Phi _{S}^{\dag }\left( x\right) \left( D^{\mu }D_{\mu}-
m^{2}\right) \Phi _{S}\left( x\right)   \label{equation883} \\
& =\int d^{4}x\,\mathcal{L}_{S} \\
\mathcal{L}_{S}& =-\left[ D^{\mu }\Phi _{S}\left( x\right) \right] ^{\dag }D_{\mu}
\Phi _{S}\left( x\right) - \Phi _{S}^{\dag }\left( x\right) m^{2} \, \Phi _{S}\left( x\right) 
\label{equation889}
\end{align}
(after integration by parts with the assumption that the boundary terms
vanish).

Since the effects of a spinor rotation cancel in (\ref{equation883}), we can
redefine $\Phi _{S}$ to have spin $0$. (If $\Phi _{S}^{(0)}$ is the value of $\Phi _{S}$ before a
rotation, and the value afterward is $U_{\mathrm{rot}} \Phi _{S}^{(0)}$, with 
$U_{\mathrm{rot}}^{\dag} U_{\mathrm{rot}} = 1$, (\ref{equation883}) is unchanged
if we replace $\Phi _{S}$ by $\Phi _{S}^{(0)}$, and then rename the fields with $\Phi _{S}^{(0)}\rightarrow
\Phi _{S}$.) $\Phi _{S}$ is then a scalar boson field (with a novel multicomponent
structure), and the theory has the usual Lorentz invariance.

Quantization follows the same prescription as for ordinary scalar fields: 
For a general complex multicomponent bosonic free field with components $\Phi_i$ we have 
\begin{align}
\pi _{i}\left( x\right)  &=\frac{\partial \mathcal{L}}{\partial \left(
\partial _{0}\Phi _{i}\left( x\right) \right) }=-\partial ^{0}\Phi_{i}^{\ast
}\left( x\right) =\partial _{0}\Phi_{i}^{\ast }\left( x\right)  \\
\pi _{i}^{\ast }\left( x\right)  &=\frac{\partial \mathcal{L}}{\partial
\left( \partial _{0}\Phi_{i}^{\ast }\left( x\right) \right) }=-\partial
^{0}\Phi_{i}\left( x\right) =\partial _{0}\Phi_{i}\left( x\right) 
\end{align}
(according to our $(-+++)$ convention for the metric tensor), so, with implied summation over $i$,
\begin{align}
\mathcal{H} &=\pi _{i}\left( x\right) \partial _{0}\Phi_{i}\left( x\right)
+\pi _{i}^{\ast }\left( x\right) \partial _{0}\Phi_{i}^{\ast }\left(
x\right) -\mathcal{L} \\
&=\partial _{0}\Phi^{\dag }\left( x\right) \partial _{0}\Phi\left( x\right)
+\partial _{k}\Phi ^{\dag } \left( x\right) \partial _{k}\Phi\left( x\right) \nonumber \\
& \hspace{1.7in}+\Phi^{\dag }\left( x\right) m^{2} \Phi\left( x\right) \;.
\end{align}

In the treatment above, $\Phi$ is a classical field. In the treatment below, to
avoid complicating the notation, $\Phi$ is taken to be the corresponding
quantum field. It is required to satisfy the usual equal-time commutation
relations: 
\begin{align}
\left[ \Phi_{i}\left( \overrightarrow{x},x^{0}\right) ,\pi _{j}\left( \overrightarrow{x}^{\prime },x^{0}\right) \right] _{-}
=i \, \mathbf{\delta }_{ij}\delta \left( \overrightarrow{x}-\overrightarrow{x}^{\prime
}\right) \;
\end{align}
with $\pi _{j}\left( \overrightarrow{x}^{\prime },x^{0}\right)
=\partial _{0}\Phi_{j}^{\ast} \left( \overrightarrow{x}^{\prime },x^{0}\right) $, and
with the other relation just being the Hermitian conjugate.

To achieve this, we follow the usual procedure, first writing $\Phi$ in terms
of the usual destruction and creation operators $c_{p}^{s}$ and $d_{p}^{s\,\dag }$, 
where $s$ distinguishes states with the same 4-momentum $p$:
\begin{align}
\Phi\left( x\right) =\int \frac{d^{3}p}{\left( 2\pi \right) ^{3}}\frac{1}{\sqrt{
2p_{0}}}\sum\limits_{s}\left( c_{p}^{s}\,U_{p}^{s}\,e^{ip\cdot
x}+d_{p}^{s\,\dag }V_{p}^{s}\,e^{-ip\cdot x}\right)
\end{align}
with $p_{0}$ on-shell. 
The particle and antiparticle states are respectively
$\sqrt{2p_{0}} c_{p}^{s\,\dag }\left\vert 0\right\rangle $ and $\sqrt{2p_{0}}
d_{p}^{s\,\dag }\left\vert 0\right\rangle $. 

We require as usual
\begin{align}
& \left[ c_{p}^{s},c_{p^{\prime }}^{s^{\prime }\,\dag }\right] _{-}=\delta
^{ss^{\prime }}\left( 2\pi \right) ^{3}\delta \left( \overrightarrow{p}- 
\overrightarrow{p}^{\prime }\right) \\
& \left[d_{p}^{s},d_{p^{ \prime }}^{s^{\prime }\,\dag }\right] _{-}=\delta
^{ss^{\prime }}\left( 2\pi \right) ^{3}\delta \left( \overrightarrow{p}- 
\overrightarrow{p}^{\prime }\right)
\end{align}
where $\overrightarrow{p}$ is the 3-momentum, with the other commutators
equaling zero. The $U_{p}^{s}$ and $V_{p}^{s}$ are orthonormal vectors
satisfying 
\begin{align}
\sum\limits_{s}U_{p}^{s}\,U_{p}^{s\,\dag }=\mathbf{1}\quad ,\quad
\sum\limits_{s}V_{p}^{s}\,V_{p}^{s\,\dag }=\mathbf{1}  \label{pol}
\end{align}
where $\mathbf{1}$ is the identity matrix. The above properties then imply
that 
\begin{align}
& \left[ \Phi_{i}\left( \overrightarrow{x},x^{0}\right) , \partial
_{0}\Phi_{j}^{\dag }\left( \overrightarrow{x}^{\prime },x^{0}\right) \right]
_{-}  \notag \\
& \hspace{0.1in} =\int \frac{d^{3}p}{\left( 2\pi \right) ^{3}}
e^{i\overrightarrow{p}\cdot \left( \overrightarrow{x}-\overrightarrow{x}^{\prime
}\right) } \frac{+ip_{0}}{ 2p_{0}}\sum\limits_{s}\left(
U_{pi}^{s}\,U_{pj}^{s\,\dag }+V_{pi}^{s}\,V_{pj}^{s\,\dag }\right)  \notag \\
& \hspace{0.3in} =i\,\mathbf{\delta }_{ij}\delta \left( \overrightarrow{x}-
\overrightarrow{ x }^{\prime }\right) \;.
\end{align}
The Hamiltonian is obtained from 
\begin{align}
H_{\Phi} &=\int d^{3}x\,\mathcal{H}\left( x\right) \\
&=\int \frac{d^{3}p}{\left( 2\pi \right) ^{3}} \frac{p_0^2 + p_0^2}{2 p_0}  \sum\limits_{s} \left(
c_{p}^{s\dag }\,\,c_{p}^{s}\,+d_{p}^{s}\,d_{p}^{s\dag }\right) \\
&=\int \frac{d^{3}p}{\left( 2\pi \right) ^{3}} \, p_{0} \sum\limits_{s} \left(
n_{p}^{s}\,+\overline{n}_{p}^{s} + 1 \right) 
\label{ham}\\
n_{p}^{s}\, &=c_{p}^{s\dag }\,\,c_{p}^{s} \; \, ,\; \, \overline{n}
_{p}^{s}=d_{p}^{s\dag }\,d_{p}^{s} \; .
\end{align}

Below we will find that each complex field $\Phi$ here should actually be separated into two real fields:
\begin{align}
\Phi = H + i H^{\prime} \; .
\label{007}
\end{align}
For each such real field the above treatment can be modified to give the same basic results with 
\begin{align}
d_{p}^{s}=c_{p}^{s} \quad , \quad V_{p}^{s} = U_{p}^{s \, \ast} 
\end{align}
and a factor of $1/2$ in (\ref{ham}) for each of the two independent fields.
Each higgson particle $H$ or $H^{\prime}$ is thus its own antiparticle.

Let us now specialize to the electroweak theory. The full gauge covariant
derivative after symmetry-breaking is~\cite{Peskin} 
\begin{align}
D_{\mu }=\partial _{\mu }& -i\frac{g}{\sqrt{2}}\left( W_{\mu
}^{+}T^{+}+W_{\mu }^{-}T^{-}\right)   \notag \\
& -i\frac{g}{\cos \theta _{w}}Z_{\mu }\left( T^{3}-\sin ^{2}\theta
_{w}\,Q\right) -ieA_{\mu }\,Q  \label{e5}
\end{align}
where $A_{\mu }$ is now the electromagnetic field. First consider the
coupling of $\Phi _S$ to $Z$, which results from the Lagrangian 
\begin{align}
& \Phi _S^{\dag }\left( \partial ^{\mu }-ig_{Z}Z^{\mu }T^{3}\right) \left(
\partial _{\mu }-ig_{Z}Z_{\mu }T^{3}\right) \Phi _S \nonumber \\
& \hspace{0.2in}=\Phi _S^{\dag }\left( \partial ^{\mu }\partial _{\mu
}-2ig_{Z}Z^{\mu }T^{3}\partial _{\mu }-\frac{1}{4}g_{Z}^{2}Z^{\mu }Z_{\mu
}\right) \Phi _S
\end{align}
(since $\left( T^{3}\right) ^{2}=1/4$), where 
$g_{Z}=g/\cos \theta _{w}$. There are no terms involving $\partial ^{\mu
}Z_{\mu }$ etc. because 
\begin{align}
\partial ^{\mu }Z_{\mu }=0\quad ,\quad \partial ^{\mu }W_{\mu }^{\pm }=0
\end{align}
follows from the equations of motion for the massive vector boson fields 
$Z_{\mu }$ and $W_{\mu }^{\pm }$. The simplest first-order interaction
relevant to a neutral particle is then 
\begin{align}
\mathcal{L}_{1}^{Z}=-2i\,g_{Z}\,Z^{\mu }\Phi _ST^{3}\partial _{\mu }\Phi _S\;.
\end{align}
For each plane-wave state $-i\,\partial _{\mu} \Phi _S=p_{\mu }
\Phi _S$, so $\Phi _S^{\dag }\,T^{3}\Phi _S=0$
implies $\mathcal{L}_{1}^{Z}=0$ . The first-order couplings involving 
$W^{\mu \pm }$ also vanish: 
\begin{align}
\mathcal{L}_{1}^{W}=-i\sqrt{2}g\Phi _S\left( W^{\mu +}T^{+}+W^{\mu -}T^{-}\right)
\partial _{\mu }\Phi _S=0\;.
\end{align}

We must then turn to the second-order interactions, 
\begin{align}
\hspace{-0.2cm}\mathcal{L}_{2}^{Z}=-\frac{g_{Z}^{2}}{4}\Phi _S^{\dag }Z^{\mu }Z_{\mu }\Phi _S \;\; , \;\;
\mathcal{L}_{2}^{W}=-\frac{g^{2}}{2}\Phi _S^{\dag }W^{\mu +}W_{\mu }^{-}\Phi _S
\end{align}
(since $\left( T^{+}\right) ^{2}=\left( T^{-}\right) ^{2}=0$ and 
$T^{+}T^{-}+T^{-}T^{+}=1$). 

Substitution of (\ref{007}) gives, in an obvious notation (for each $S$)
\begin{align}
\hspace{-0.2cm}\mathcal{L}_{H}^{Z} &=-\frac{g_{Z}^{2}}{4}H^{\dag }Z^{\mu }Z_{\mu }H \;\; , \;\;
\mathcal{L}_{H}^{W}=-\frac{g^{2}}{2}H^{\dag }W^{\mu +}W_{\mu }^{-}H 
\end{align}
and the same for $H^{\prime}$. The lower-mass field is chosen to be $H$. 

The lowest-mass of all higgson fields 
is called $H^{0}$, and its gauge interactions are given by
\begin{align}
\mathcal{L}_{0}^{Z}=-\frac{g_{Z}^{2}}{4}H^{0\dag }Z^{\mu }Z_{\mu
}H^{0}\;\;,\;\;\mathcal{L}_{0}^{W}=-\frac{g^{2}}{2}H^{0\dag }W^{\mu +}W_{\mu
}^{-}H^{0}\;.
\label{e8}
\end{align}
The particle $H^{0}$ is stable because the interactions of (\ref{e8}) (as well as the 
undetermined Higgs coupling) do not permit it to decay: If there is a single
initial $H^{0}$, the final set of products must also contain $H^{0}$, since
it is coupled only to itself and either two gauge bosons or the Higgs boson. 
$H^{0}$ can lose energy by radiating other particles, but it cannot decay.
In this context, one might note that only the final (physical)
Lorentz-invariant fields and action determine what
processes are allowed, even if those processes are virtual.

\begin{figure}[tbp]
\begin{center}
\resizebox{0.45\columnwidth}{!}{
\includegraphics{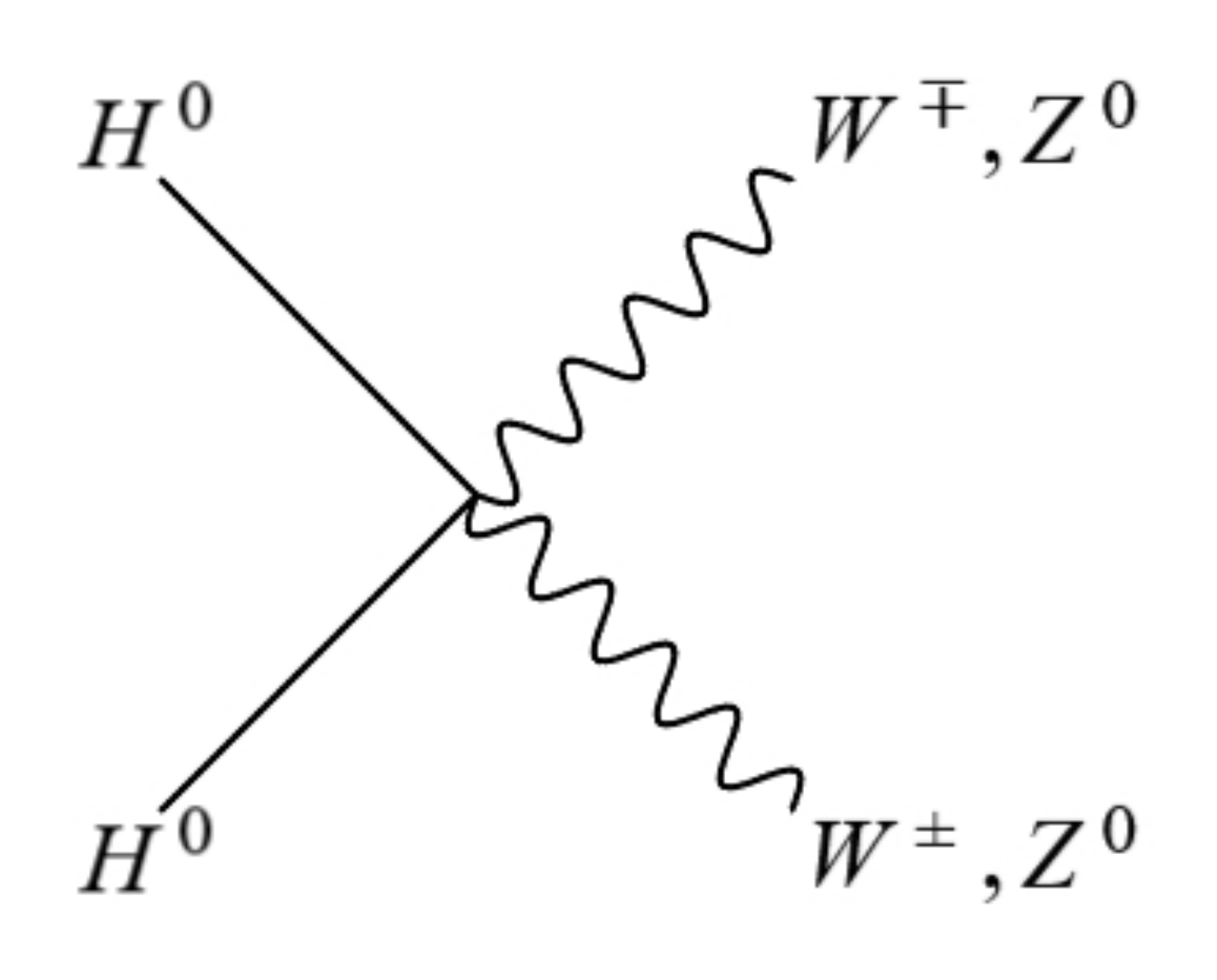}}
\end{center}
\caption{Annihilation into real W or Z boson pair, for $m_H>$ 90 GeV. These
processes would produce a relic abundance of dark matter that is more than
an order of magnitude too low to agree with the observed dark matter density.}
\label{annihilation-WZ}
\end{figure}

\begin{figure}[tbp]
\begin{center}
\resizebox{0.49\columnwidth}{!}{
\includegraphics{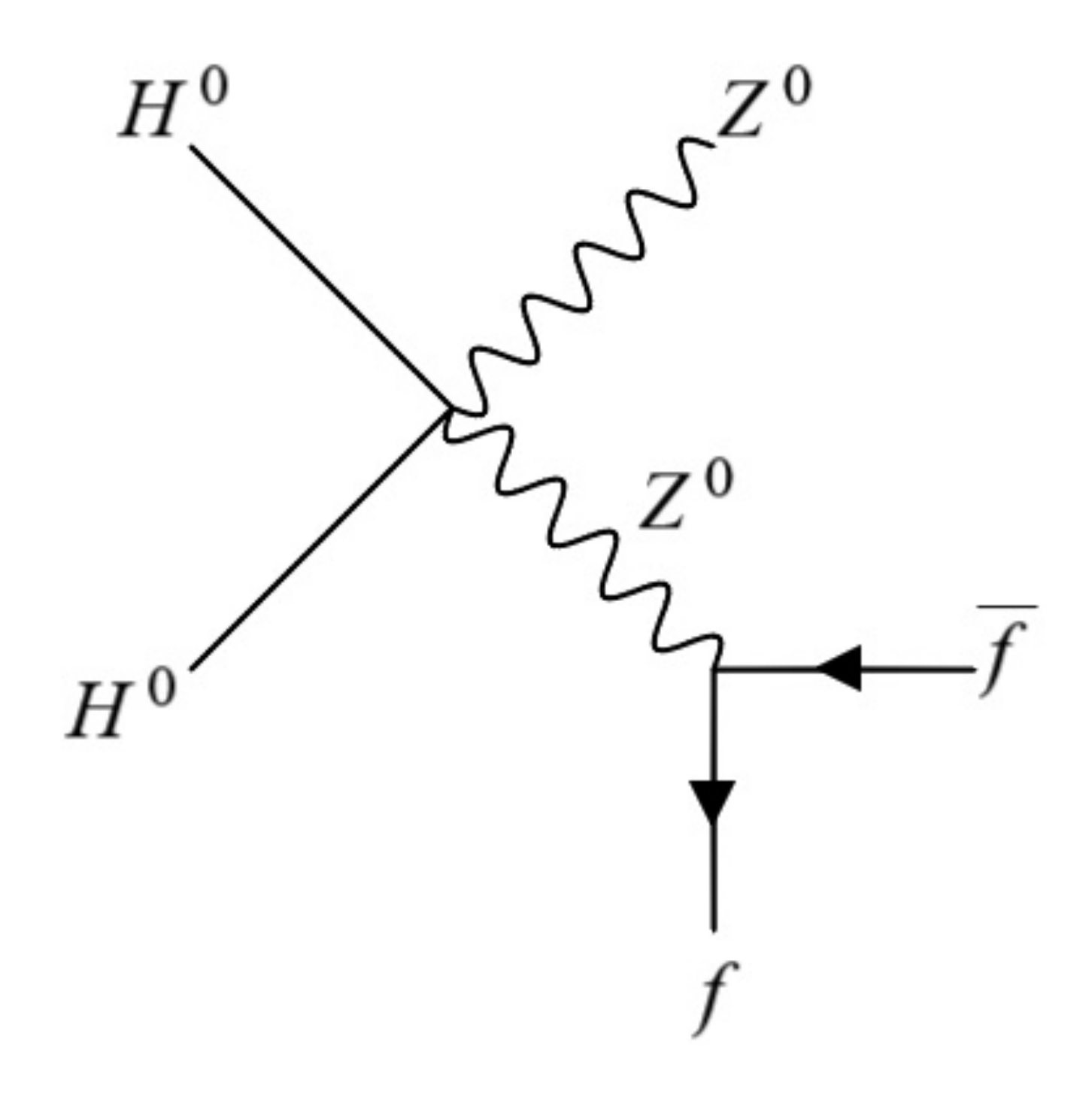}} 
\resizebox{0.49\columnwidth}{!}{
\includegraphics{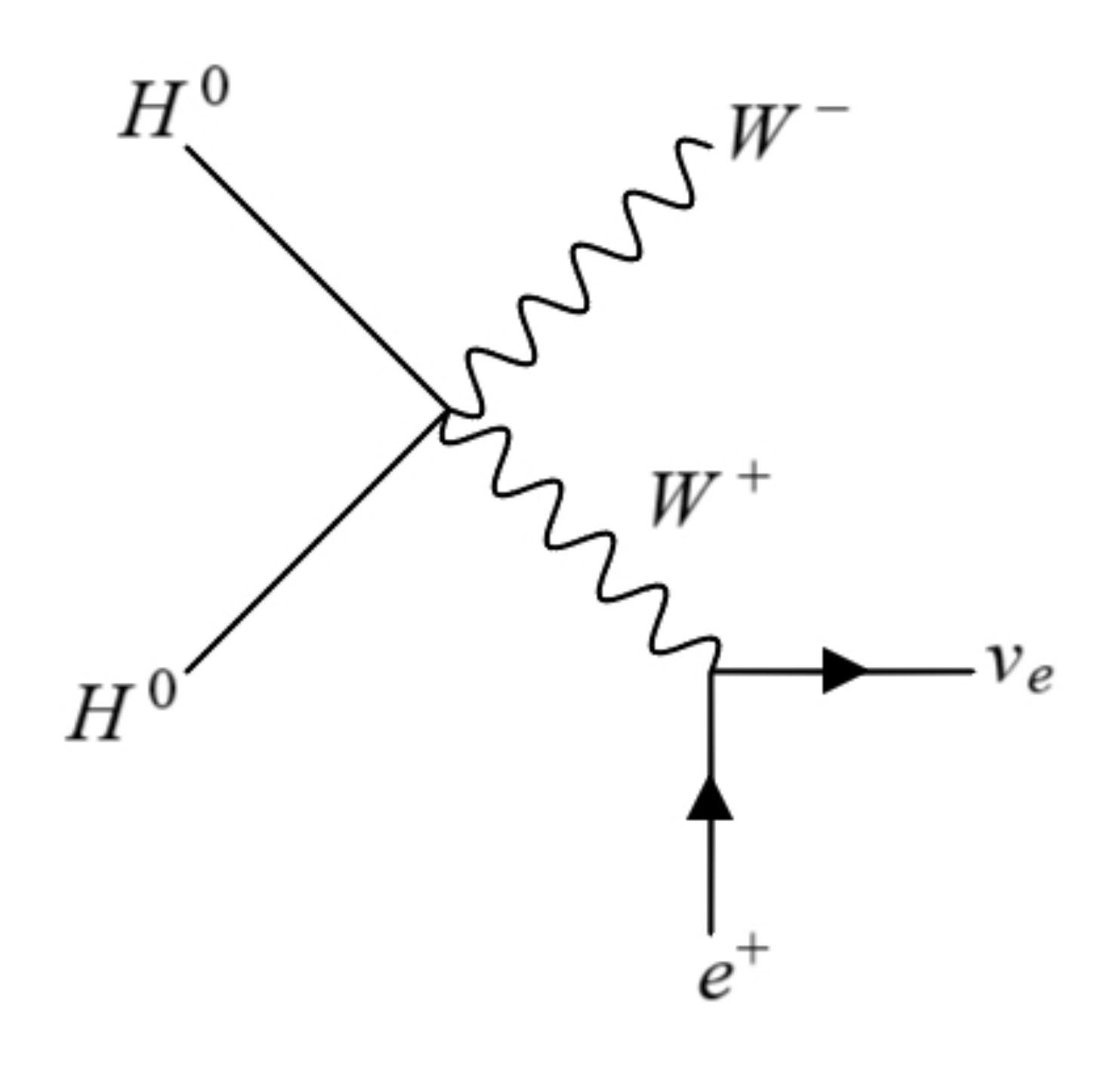}}
\end{center}
\caption{Left panel: Annihilation into one real and one virtual Z boson, for
90 GeV $>m_H>50$ GeV. There are 11 such processes for fermion-antifermion
pairs $f\bar{f}$ (6 for leptons and 5 for the quarks with sufficiently low
masses). Right panel: Example of annihilation into one real and one virtual
W boson for 80 GeV $>m_H>43$ GeV. There are 18 such processes for
fermion-antifermion pairs (6 for leptons and 12 for the quarks with
sufficiently low masses). The total annihilation cross-section from these
processes will be consistent with the observed relic abundance for $m_H \sim 72$ GeV.}
\label{WZ}
\end{figure}

We have performed approximate calculations of the annihilation
cross-sections for the lowest-order processes involving the second-order
couplings of (\ref{e8}), shown in Figures~\ref{annihilation-WZ} and \ref{WZ}, using
standard methods~\cite{Peskin, Cheng}. The only new (and trivial) feature is
the sums over higgson states using (\ref{pol}). 
For the processes of Fig. \ref{annihilation-WZ}, we make the approximation
that the $W$, $Z$, and $H$ masses are nearly equal ($\sim$~80-100 GeV.). We
find that the total annihilation cross-section is more than an order of
magnitude too large for $m_Z > m_H > m_W$, and about a factor of 2 larger
still for $m_H > m_Z$. (Without this approximation, the cross-sections would
be even larger.) I.e., we obtain 
\begin{align}
\langle \sigma_{ann} v \rangle \gg \langle \sigma v \rangle_S \quad \
\mathrm{for} \quad m_H>m_W\approx 80 \, \mathrm{GeV}  \label{sigma-S}
\end{align}
where $\langle \sigma v \rangle_S = 2.2 \times 10^{-26} \, \mathrm{cm}^3/\mathrm{s}$
is the benchmark value obtained by Steigman et al.~\cite{Steigman} for a
WIMP with mass above 10 GeV that is its own antiparticle, if the relic dark
matter density is to agree with astronomical observations.

For the processes of Fig.~\ref{WZ}, we make the approximation of neglecting
the masses of the fermions (which are all small compared to $m_Z$, $m_W$,
and $m_H$). There are 29 processes involving fermion-antifermion pairs. If 
$m_H$ is well below $m_W$, the total cross-section is more than an order of
magnitude too small (compared again to $\langle \sigma v \rangle_S)$.

As $m_{H}$ approaches $m_{W}$ from below, however, there is resonance-like 
behavior involving the $W$ propagator, and there is some value of $m_{H}$
(not precisely determined here because of the approximations described
above) such that 
\begin{align}
\langle \sigma _{ann}v\rangle \sim \langle \sigma v\rangle _{S}\quad 
\mathrm{with}\quad m_{H}\sim 72\,\mathrm{GeV}\;.
\end{align}

This annihilation cross-section is consistent with the limits set by
observation of gamma-ray emissions from dwarf spheroidal galaxies by
Fermi-LAT~\cite{Leane-1,gammas-okay,Leane-2}, if the annihilation is treated
generally rather than simplistically assumed to proceed through a single
channel. For the present particle, there are 29 annihilation processes,
represented by Fig.~\ref{WZ}.

This cross-section and mass are also consistent with analyses of the gamma
ray excess from the Galactic center observed by Fermi-LAT~\cite{Goodenough,Fermi,Fermi-GCE,Leane,Cuoco2}, 
and with analyses of the antiprotons observed by AMS~\cite{Cuoco2,Cuoco,Cui,AMS-1,AMS-2}, 
which independently have been interpreted as potential evidence of dark
matter annihilation -- although there are, of course, competing interpretations 
based on backgrounds and alternative statistical approaches. The inferred values of the particle mass and
annihilation cross-section are in fact remarkably similar to those obtained
here; see e.g. the abstracts of Refs.~\cite{Fermi-GCE} and \cite{AMS-1}, and
Fig. 12 of Ref.~\cite{AMS-2}.

The predictions of the present theory within this context are very similar to those for the IDM if the 
masses of $A^0_I$ and $H^{\pm}_I$ are well separated from that of $H^0_I$.
A very detailed analysis of the Fermi-LAT
gamma-ray data, and its comparison with IDM predictions, has been given in \cite{Eiteneuer}.
For annihilations of a dark matter WIMP with a mass of $\sim 72$ GeV  
the basic qualitative conclusions are the same as above.

\begin{figure}[tbp]
\centering
\resizebox{0.45\columnwidth}{!}{
\includegraphics{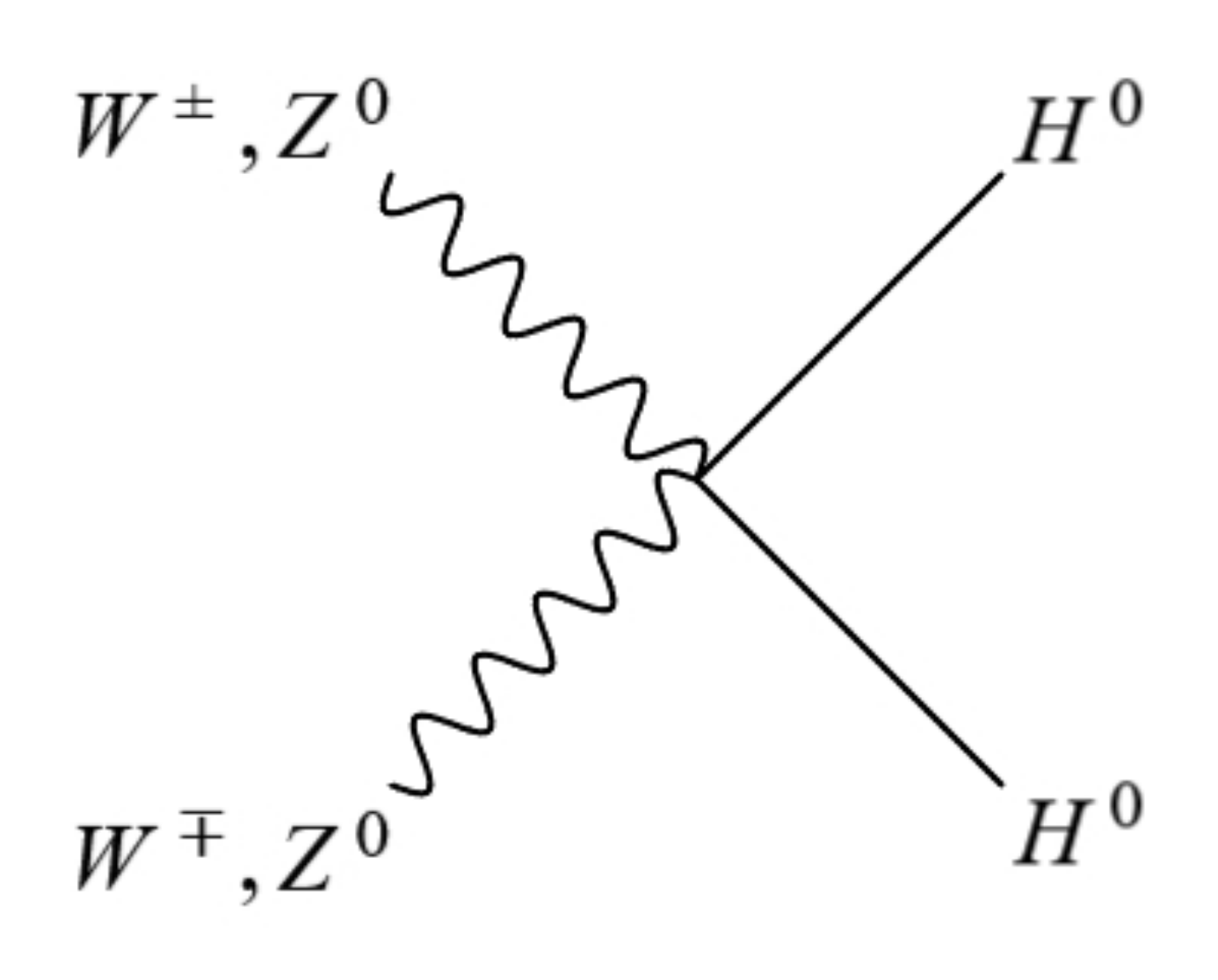}}
\caption{Collider creation through vector boson fusion.}
\label{creation-fusion}
\end{figure}

Collider detection of dark matter particles often focuses on creation
through the $Z$, Higgs, or some hypothetical new mediator~\cite{collider}.
For the present particle there is no first-order coupling to the $Z$ and
there are no exotic new mediators, so for this kind of process only the
Higgs portal remains a possibility. CMS and ATLAS have independently placed
upper limits on the branching ratio for invisible Higgs decays to particles
with a total mass of $<125$~GeV~\cite{CMS-2019,ATLAS-2019}. 
The present particle may have a small Higgs coupling, however, and 
the total mass of a pair should be $\sim $ 145 GeV. 
It appears that the present particle is also consistent with other
collider-detection limits, and that the best possibility for observation in
a collider experiment is the process depicted in Fig.~\ref{creation-fusion}.
The predicted signature for collider detection is then $\gtrsim$~145~GeV of
missing transverse energy resulting from vector boson fusion (VBF). 
In the present context this is a weak process, but the results of
Refs. \cite{Dutta} and \cite{Dercks-2019} (for the IDM), and \cite{Bishara-2017} and \cite{Dreyer-2020}  (for double Higgs production) 
suggest that observation of this process may be barely possible with a sustained
run of the high-luminosity LHC, if an integrated luminosity of up to 3000 fb$^{-1}$ can be achieved. 
(Definitive studies may have to await a 100 TeV collider.) If there is a contribution 
from Higgs coupling, of course, the signal for creation of these particles will be stronger.
\begin{figure}[tbp]
\begin{center}
\resizebox{0.49\columnwidth}{!}{
\includegraphics{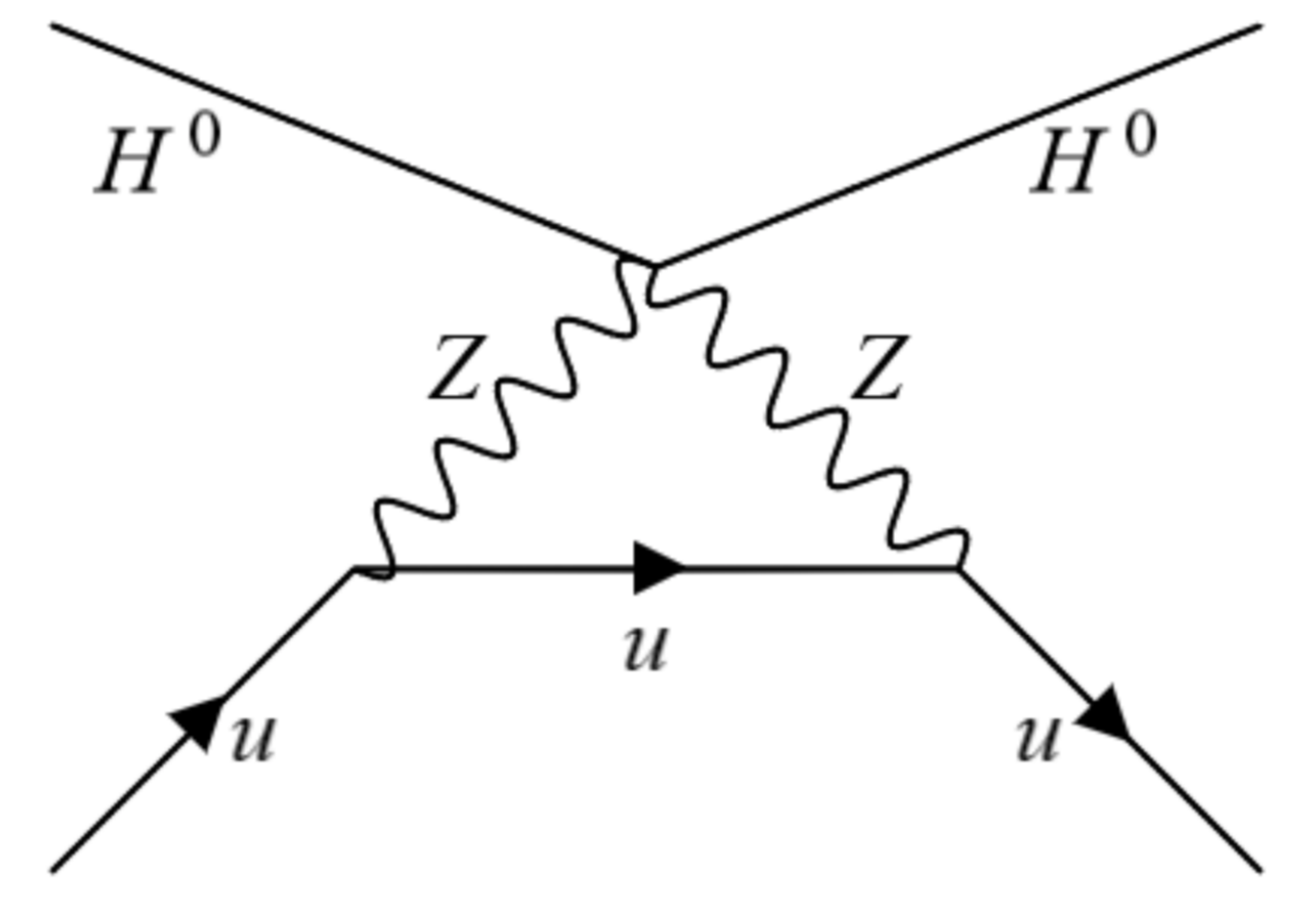}} 
\resizebox{0.49\columnwidth}{!}{
\includegraphics{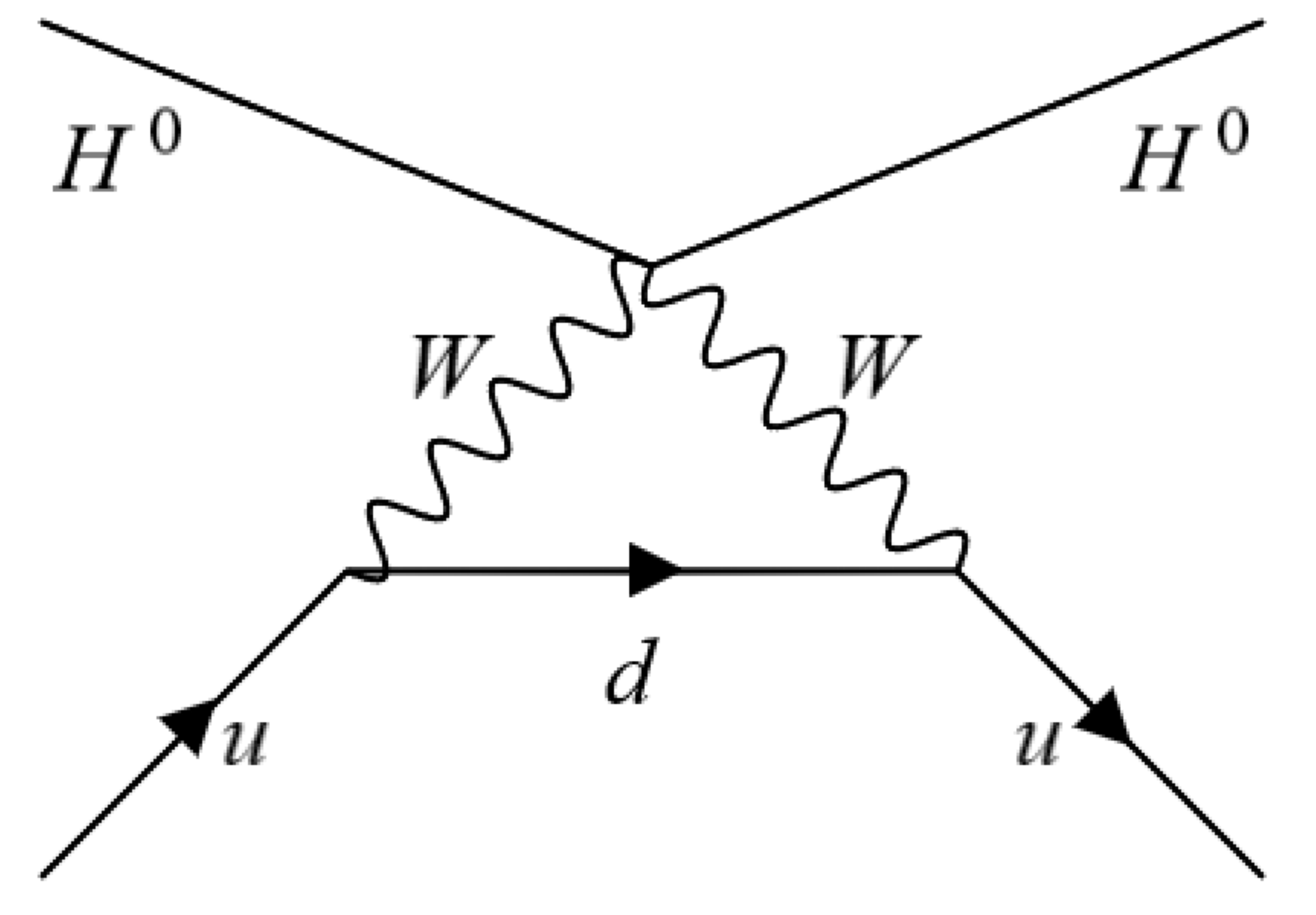}}
\end{center}
\caption{Scattering through loop processes, with a low but potentially accessible cross-section.}
\label{one-loop}
\end{figure}

The best hope for direct detection appears to be the one-loop processes of 
Fig.~\ref{one-loop}, which are the same as for the IDM in Fig.~1 of  \cite{Klasen-2013}. 
The results in Figs. 2 and 3 of that paper suggest that 
this mechanism -- scattering with an exchange of two Z or W bosons -- has 
a cross-section not far below $10^{-11}$ pb for the present particle -- perhaps barely 
within reach of sustained runs for the next generation of direct-detection experiments. Definitive studies 
would probably require even greater sensitivity, but with a cross-section still potentially 
attainable (and above the neutrino floor).
Of course, there is again the possibility of some enhancement from Higgs exchange.

The present scenario is quite amenable to being tested by experiment and observation. For
example, if the positron excess observed by AMS~\cite{Hooper-positron} (and
other experiments) had demonstrated that the dominant dark matter particle
has a mass of roughly 800 GeV or above, the much lower mass in the present
scenario would have been disconfirmed. But the Planck observations have
instead disconfirmed this interpretation~\cite{Planck}. 
On the other hand, the  interpretation of a lower-mass dark-matter signal in the analyses cited
above is quite consistent with the Planck data, and in addition is just as
expected for thermal production. 

We conclude with a broad comment: The behavior of spin 1/2 fermions and spin
1 gauge bosons has turned out to be far richer than originally envisioned
(dating from the discovery of the electron in 1897 and the proposal of the
photon in 1905), so one might anticipate further richness connected to the
recent discovery of a spin 0 boson. In the present theory this boson
represents the lowest-energy amplitude excitation in an extended Higgs
sector.

\acknowledgments
We are greatly indebted to an anonymous reviewer, whose helpful suggestions have led to 
many major improvements in the paper. The relevance of the inert doublet model, the 
importance of the processes in Fig.~\ref{one-loop}, the need for calculations of the processes in 
Figs.~\ref{annihilation-WZ} and \ref{WZ}, and the need to cite early analyses of gamma-ray 
and antiproton data are a few of the suggestions that have resulted in large positive changes 
in the paper.

\end{document}